\documentclass[twocolumn,english,longbibliography]{revtex4-2}
\usepackage[T1]{fontenc}
\usepackage[latin9]{inputenc}
\usepackage[a4paper]{geometry}
\usepackage{babel}
\usepackage{array}
\usepackage{booktabs}
\usepackage{mathrsfs}
\usepackage{amsmath}
\usepackage{stmaryrd}
\usepackage{graphicx}
\usepackage[unicode=true,pdfusetitle,
 bookmarks=true,bookmarksnumbered=false,bookmarksopen=true,bookmarksopenlevel=2,
 breaklinks=false,pdfborder={0 0 1},backref=false,colorlinks=false]
 {hyperref}
\hypersetup{
 colorlinks=true,allcolors=blue}

\makeatletter

\providecommand{\tabularnewline}{\\}

\@ifundefined{date}{}{\date{}}
\usepackage{color,soul}
\usepackage{upgreek}
\date{\today}
\newcommand{\header}[1]{}

\makeatother

\begin{document}
\title{Inverse design of focused vector beams\\ for mode excitation in optical
nanoantennas}
\author{Xiaorun Zang}
\email{xiaorun.zang@aalto.fi}

\altaffiliation{Currently at Department of Applied Physics, Aalto University, P.O Box 13500, FI-00076 Aalto, Finland}

\affiliation{Institute of Photonics, University of Eastern Finland, P.O. Box 111,
FI-80101 Joensuu, Finland}
\author{Ari T. Friberg}
\affiliation{Institute of Photonics, University of Eastern Finland, P.O. Box 111,
FI-80101 Joensuu, Finland}
\author{Tero Setälä}
\affiliation{Institute of Photonics, University of Eastern Finland, P.O. Box 111,
FI-80101 Joensuu, Finland}
\author{Jari Turunen}
\affiliation{Institute of Photonics, University of Eastern Finland, P.O. Box 111,
FI-80101 Joensuu, Finland}
\begin{abstract}
We propose a free-space, inverse design of nanostructure's effective
mode-matching fields via a backward propagation of tightly focused
vector beams to the pupil plane of an aplanatic system of high numerical
aperture. First, we study the nanostructure's eigenmodes without considering
any excitation fields and then extract the modal near fields in the
focal plane. Each modal field is then taken as the desired focal field,
the band-limited waves of which are backward propagated to the pupil
plane via a reversal of the Richards--Wolf vector diffraction formula.
The pupil fields can be designed to be genuinely paraxial by associating
the longitudinal electric/magnetic field component with the radial
one on the reference sphere. The inversely designed pupil field in
turn is propagated forwardly into the focal region to generate the
designed focal field, whose distribution over the nanostructure's
surface is used to evaluate the overlap between the designed focal
field and the modal fields, i.e., the modal expansion coefficients.
Studies for a silicon nanodisk monomer, dimer, and tetramer demonstrate
the ability of our inverse approach to design the necessary tightly
focused vector field that can effectively and exclusively match a
certain eigenmode of interest. Compared with the forward beam-shaping
method, the inverse design approach tends to yield quantitatively
more precise mode-matching field profiles. This work can have a significant
impact on optical applications that rely on controllable and tunable
mode excitation and light scattering.
\end{abstract}
\maketitle

\section{Introduction}

Strong local field in plasmonic or high-index dielectric nanostructures
is beneficial for boosting a variety of optical effects, via an enhanced
interaction between the strong light field and nanostructures. In
contrast to plasmonic systems, high-index dielectric nanoparticles,
such as silicon nanodisks \citep{staudeTailoringDirectionalScattering2013,chongPolarizationindependentSiliconMetadevices2015,baranovAnapoleenhancedIntrinsicRaman2018,zhaoSwitchableChiropticalHotspots2019},
do not suffer from strong optical losses but still sustain deep sub-wavelength
field confinements \citep{kuznetsovOpticallyResonantDielectric2016}.
Moreover, all-dielectric nanostructures have been demonstrated to
be versatile semiconductor-compatible nanophotonic devices for directional
scattering \citep{staudeTailoringDirectionalScattering2013}, near-field
mapping of optical modes \citep{habteyesNearfieldMappingOptical2014},
tailoring second-harmonic \citep{cambiassoBridgingGapDielectric2017,camacho-moralesResonantHarmonicGeneration2019,sautterTailoringSecondHarmonicEmission2019}
and third-harmonic \citep{shcherbakovEnhancedThirdharmonicGeneration2014}
generations, anapole-enhanced Raman scattering \citep{baranovAnapoleenhancedIntrinsicRaman2018},
etc. With recent advances, all-dielectric nanostructures are promising
to form basic building blocks that complement or even replace the
plasmonic counterparts in modern nanophotonics \citep{kuznetsovOpticallyResonantDielectric2016}.

At certain frequencies the local field enhancement achieves maximum
values (i.e., peaks), which is a phenomenon described as the optical
resonance \citep{maierPlasmonicsLocalizationGuiding2005,kuznetsovOpticallyResonantDielectric2016}.
It is known that the resonance response can be tuned by varying size,
shape, and dielectric environment of the nanostructures \citep{kellyOpticalPropertiesMetal2003}.
The resonance is commonly analyzed with respect to a certain incident
light field, such as a plane wave used in previous studies \citep{staudeTailoringDirectionalScattering2013,morenoAnalysisSpectralBehavior2013,baranovAnapoleenhancedIntrinsicRaman2018},
which, however, may not probe all resonances \citep{makitaloModesResonancesPlasmonic2014}.
Recently, it has been shown that vector beams \citep{zhanCylindricalVectorBeams2009,rosales-guzmanReviewComplexVector2018}
with spatially varying polarization distributions can reveal otherwise
hidden resonances, such as the dark resonance detected by the radially
or azimuthally polarized cylindrical vector beams \citep{sancho-parramonDarkModesFano2012,gomezDarkSidePlasmonics2013,yanaiNearandFarfieldProperties2014},
or the hybrid resonances by higher-order vector beams possessing helical
phase distributions \citep{reichSelectionRulesStructured2020,zang*EfficientHybridmodeExcitation2021}.

The optical resonance response in nanostructures is associated with
the excitation of one or more mode(s) \citep{makitaloModesResonancesPlasmonic2014}
which, strictly speaking, are independent of any external excitation
field and hence are intrinsically determined by the optical properties
of the nanostructure and surrounding media. The resonance frequencies
are generally isolated poles in the complex frequency plane, a characteristic
of quasi-normal modes \citep{lalanneLightInteractionPhotonic2018}.
At a fixed real-valued frequency, however, a set of discrete eigenmodes
\citep{makitaloModesResonancesPlasmonic2014,powellResonantDynamicsArbitrarily2014}
can be defined. Modes in nanostructures have been investigated by
various approaches including the electrostatic method \citep{fredkinResonantBehaviorDielectric2003,mayergoyzElectrostaticPlasmonResonances2005,gomezSymmetryEffectsOptical2010},
and several rigorous full-wave analyses based on the Fourier modal
method \citep{weissDerivationPlasmonicResonances2011,bykovUseAperiodicFourier2017,bykovNumericalMethodsCalculating2013},
transfer matrix method \citep{suryadharmaStudyingPlasmonicResonance2017},
finite element method \citep{guo3dimensionalEigenmodalAnalysis2012,lalanneLightInteractionPhotonic2018},
Green's tensor method \citep{paulusLightPropagationScattering2001},
and boundary element method (BEM) based on the surface integral equations
(SIEs) \citep{makitaloModesResonancesPlasmonic2014,powellResonantDynamicsArbitrarily2014,bernasconiModeAnalysisSecondharmonic2016,powellInterferenceModesAlldielectric2017},
etc. Among them, BEM has been proven to be efficient particularly
in modelling the interaction between focused vector field and nanostructures
\citep{bautistaSecondharmonicGenerationImaging2012,bautistaSecondharmonicGenerationImaging2015,zang*EfficientHybridmodeExcitation2021},
because the numerical discretization as well as the corresponding
focal fields evaluation are reduced to two dimension (2D) for a general
three dimensional (3D) problem.

The excitation efficiency of an eigenmode scales with the overlap
between the modal and the applied incident field profiles, as well
as the corresponding eigenvalue \citep{makitaloModesResonancesPlasmonic2014}.
An eigenmode's near field in optical nanoantennas is vectorial and
localized such that it may vary considerably within a sub-wavelength
scale, which is a fact that poses a challenge to the choice of an
excitation field with an effective mode-matching profile. A free-space
approach to obtain such a mode-matching vector field is to shape a
focal field that is to match with the modal field, by focusing a paraxial
vector beam in an aplanatic system with high numerical aperture (NA).
A forward design approach has been shown to yield spatially varying
field distributions for efficient hybrid mode excitations \citep{zang*EfficientHybridmodeExcitation2021},
but a forwardly designed focal field is limited to a qualitative mode
matching due to energy exchanges between the radial and azimuthal
field components in the focusing process, and then a deviation of
its field profile from the designed pupil field's. Alternatively,
an inverse design is versatile in shaping focal fields that potentially
bring more refined and effective mode-matching fields.

In the previous work of inverse problem in high-NA aplanatic system,
the main interest has been focused on the focal field total intensity
or a single component from which the complex pupil field is retrieved.
For instance, the depth of focus can be optimized after determining
the abberration and amplitude functions in high-NA imaging system
\citep{kantSuperresolutionIncreasedDepth2000}, or even retrieving
the birefringence from four focal field intensity distributions generated
by pupil fields of different polarization states \citep{braatExtendedNijboerZernike2005}.
A focal field with a null longitudinal component is generated by azimuthally
polarized light \citep{foremanInversionDebyeWolfDiffraction2008}.
Needle-, tube-, and bubble-shaped focal field intensity distributions
\citep{jahnSolvingInverseProblem2013}, perfect polarization vortex
focal field \citep{chenReverseEngineeringApproach2016}, as well as
focal field of arbitrary homogeneous polarization \citep{ruiSynthesisFocusedBeam2016}
are also investigated. However, a mode-matching focal field generally
involves more than one or all field component(s) to be effectively
matched with both the amplitude and phase distributions of the modal
field, which has not been considered in previous work.

In this work, we present an inverse design of the desired focal field
for effective mode excitation in optical nanoantennas by taking into
account all modal field components and the inherent dependencies of
all electric and magnetic field components in the designed beam-like
pupil, reference, and focal fields. Independent of any excitation
field, the eigenmodes are fully determined by the nanostructure's
optical properties themselves and thus they are studied via a BEM
mode solver implemented through the M\"uller formulation of the SIEs
\citep{yla-oijalaWellconditionedMullerFormulation2005}. The modal
fields in the focal plane are evaluated via the Stratton---Chu surface
integral \citep{strattonDiffractionTheoryElectromagnetic1939}, and
the extracted focal fields are propagated back to the pupil plane
by reversing the tight focusing process. The best possible incident
paraxial beams before focusing are then obtained from solving an inverse
problem governed by the Richards--Wolf vector diffraction formula.

This work is organized as follows. In Sec.~\ref{sec:eigenmodes_formulation},
we study the eigenmodes in nanostructures, where we review the SIEs,
describe the eigenvalue problem based on the M\"uller formulation,
and analyze the eigenmodes in silicon nanodisk monomer, dimer, and
tetramer. In Sec.~\ref{sec:inverse-focusing}, we describe the forward
and backward tight focusing processes which are governed by the Richards--Wolf
vector diffraction formula. The degrees of freedom and strategies
in our inverse design are then discussed through a systematical study
of the mode-matching field design in a silicon nanodisk monomer, dimer,
and tetramer. In Sec.~\ref{sec:excitation_coeff}, we analyze the
quality of the designed focal field in terms of modal expansion coefficients,
i.e., the overlaps between the designed focal field and all the first
twelve eigenmodes considered in each nanodisk oligomer. At last, conclusions
and perspectives are given in Sec.~\ref{sec:conclusion}.

\section{Eigenmodes in nanostructures\label{sec:eigenmodes_formulation}}

\subsection{Surface integral equations}

\begin{figure}[tb]
\includegraphics{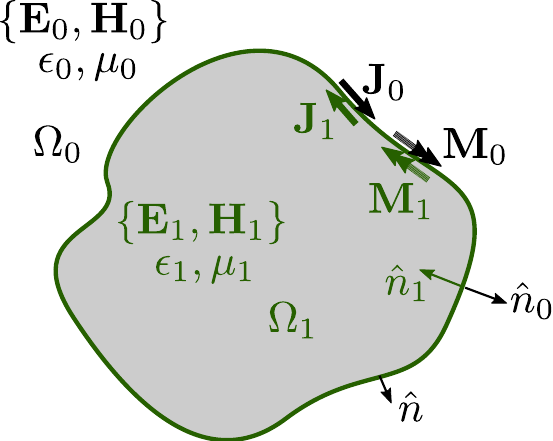}\caption{An arbitrarily-shaped 3D scatterer in free space. The scatterer that
occupies domain $\Omega_{1}$ has permittivity $\epsilon_{1}$ and
permeability $\mu_{1}$, whereas domain $\Omega_{0}$ is free space
with optical properties are $\epsilon_{0}$ and $\mu_{0}$. The electric
and magnetic fields are $\{\mathbf{E}_{0},\mathbf{H}_{0}\}$ in the
free space and $\{\mathbf{E}_{1},\mathbf{H}_{1}\}$ in the scatterer.
Two domains' surfaces are correspondingly denoted by $\partial\Omega_{0}$
and $\partial\Omega_{1}$, and their directions are determined by
normal vectors $\hat{n}_{0}=\hat{n}$ and $\hat{n}_{1}=-\hat{n}$.
Equivalent electric and magnetic surface current densities $\{\mathbf{J}_{0},\mathbf{M}_{0}\}$
and $\{\mathbf{J}_{1},\mathbf{M}_{1}\}$ are mathematically introduced
on $\partial\Omega_{0}$ and $\partial\Omega_{1}$, respectively.}
\label{fig:scatter}
\end{figure}

We first describe the eigenmode problem in optical nanostructures
using the SIE formulation. Without loss of generality, the SIE formulation
is reviewed for the case of a single arbitrarily shaped 3D nanostructure,
since it is readily extended to more general cases of nano-object
ensembles. As shown in Fig.~\ref{fig:scatter}, the nano-object occupying
domain $\Omega_{1}$ contains a homogeneous and isotropic medium with
permittivity $\epsilon_{1}$ and permeability $\mu_{1}$. The free
space in domain $\Omega_{0}$ has optical properties $\epsilon_{0}$
and $\mu_{0}$. The monochromatic electric and magnetic fields of
harmonic time dependence $e^{-i\omega t}$ in the scatterer and free
space are denoted by $\{\mathbf{E}_{1},\mathbf{H}_{1}\}$ and $\{\mathbf{E}_{0},\mathbf{H}_{0}\}$,
respectively, where $\omega$ is the angular frequency.

Applying the field equivalence principle \citep{schelkunoffEquivalenceTheoremsElectromagnetics1936,strattonDiffractionTheoryElectromagnetic1939,chewIntegralEquationMethods2009,orfanidisElectromagneticWavesAntennas2016,jinTheoryComputationElectromagnetic2015}
to the scatterer, we are interested in the fields $\{\mathbf{E}_{1},\mathbf{H}_{1}\}$
in the domain $\Omega_{1}$ and thus the fields elsewhere are not
of interest. In this regard, the fields $\{\mathbf{E}_{0},\mathbf{H}_{0}\}$
in the domain $\Omega_{0}$ can take any values while the fields in
$\Omega_{1}$ are kept fixed, which in turn implies that the tangential
fields over the surface $\partial\Omega_{1}$ of the domain $\Omega_{1}$
should be unchanged, according to the uniqueness theorem \citep{collinFieldTheoryGuided1991,kongElectromagneticWaveTheory2008,jinTheoryComputationElectromagnetic2015}.
In the equivalent problem for the scatterer, making the fields in
$\Omega_{0}$ trivial $\{\mathbf{E}_{0}=0,\mathbf{H}_{0}=0\}$ leads
to an arbitrariness in the $\epsilon_{0}$ and $\mu_{0}$ values,
and a convenient choice is to take the same values as in the scatterer
with $\epsilon_{0}=\epsilon_{1}$ and $\mu_{0}=\mu_{1}$, i.e., the
full space medium is homogeneous and isotropic. However, to maintain
the tangential fields over $\partial\Omega_{1}$, equivalent electric
and magnetic surface current densities must be introduced mathematically
\begin{align}
\mathbf{J}_{1} & =\hat{n}_{1}\times\mathbf{H}_{1},\\
\mathbf{M}_{1} & =-\hat{n}_{1}\times\mathbf{E}_{1}.
\end{align}
Applying the Stratton--Chu equation \citep{strattonDiffractionTheoryElectromagnetic1939}
in the equivalent problem for the scatterer, the fields of interest
at a point in the scatterer ($\mathbf{r}\in\Omega_{1}$) can be evaluated
by
\begin{align}
\mathbf{E}_{1}(\mathbf{r}) & =i\omega\mu_{1}\iint_{\partial\Omega_{1}}\left[\mathbf{J}_{1}\left(\mathbf{r}'\right)G_{1}\left(\mathbf{r},\mathbf{r}'\right)\right]\mathrm{d}S'\nonumber \\
 & -\frac{1}{i\omega\epsilon_{1}}\iint_{\partial\Omega_{1}}\left[\nabla'\cdot\mathbf{J}_{1}\left(\mathbf{r}'\right)\nabla G_{1}\left(\mathbf{r},\mathbf{r}'\right)\right]\mathrm{d}S'\nonumber \\
 & +\iint_{\partial\Omega_{1}}\left[\mathbf{M}_{1}\left(\mathbf{r}'\right)\times\nabla G_{1}\left(\mathbf{r},\mathbf{r}'\right)\right]\mathrm{d}S',\label{eq:SIE_E1}\\
\mathbf{H}_{1}(\mathbf{r}) & =-\iint_{\partial\Omega_{1}}\left[\mathbf{J}_{1}\left(\mathbf{r}'\right)\times\nabla G_{1}\left(\mathbf{r},\mathbf{r}'\right)\right]\mathrm{d}S'\nonumber \\
 & +i\omega\epsilon_{1}\iint_{\partial\Omega_{1}}\left[\mathbf{M}_{1}\left(\mathbf{r}'\right)G_{1}\left(\mathbf{r},\mathbf{r}'\right)\right]\mathrm{d}S'\nonumber \\
 & -\frac{1}{i\omega\mu_{1}}\iint_{\partial\Omega_{1}}\left[\nabla'\cdot\mathbf{M}_{1}\left(\mathbf{r}'\right)\nabla G_{1}\left(\mathbf{r},\mathbf{r}'\right)\right]\mathrm{d}S',\label{eq:SIE_H1}
\end{align}
where $G_{1}(\mathbf{r},\mathbf{r}')=\exp(ik_{1}|\mathbf{r}-\mathbf{r}'|)/(4\pi|\mathbf{r}-\mathbf{r}'|)$
is Green's function in the whole space of the equivalent problem,
and $k_{1}$ is the wave number with $k_{1}^{2}=\epsilon_{1}\mu_{1}\omega^{2}$.

In the equivalent problem for the free space, the fields at a point
$\mathbf{r}\in\Omega_{0}$ can be evaluated similarly by
\begin{align}
\mathbf{E}_{0}(\mathbf{r}) & =\mathbf{E}_{0}^{(i)}(\mathbf{r})+i\omega\mu_{0}\iint_{\partial\Omega_{0}}\left[\mathbf{J}_{0}\left(\mathbf{r}'\right)G_{0}\left(\mathbf{r},\mathbf{r}'\right)\right]\mathrm{d}S'\nonumber \\
 & -\frac{1}{i\omega\epsilon_{0}}\iint_{\partial\Omega_{0}}\left[\nabla'\cdot\mathbf{J}_{0}\left(\mathbf{r}'\right)\nabla G_{0}\left(\mathbf{r},\mathbf{r}'\right)\right]\mathrm{d}S'\nonumber \\
 & +\iint_{\partial\Omega_{0}}\left[\mathbf{M}_{0}\left(\mathbf{r}'\right)\times\nabla G_{0}\left(\mathbf{r},\mathbf{r}'\right)\right]\mathrm{d}S',\label{eq:SIE_E0}\\
\mathbf{H}_{0}(\mathbf{r}) & =\mathbf{H}_{0}^{(i)}(\mathbf{r})-\iint_{\partial\Omega_{0}}\left[\mathbf{J}_{0}\left(\mathbf{r}'\right)\times\nabla G_{0}\left(\mathbf{r},\mathbf{r}'\right)\right]\mathrm{d}S'\nonumber \\
 & +i\omega\epsilon_{0}\iint_{\partial\Omega_{0}}\left[\mathbf{M}_{0}\left(\mathbf{r}'\right)G_{0}\left(\mathbf{r},\mathbf{r}'\right)\right]\mathrm{d}S'\nonumber \\
 & -\frac{1}{i\omega\mu_{0}}\iint_{\partial\Omega_{0}}\left[\nabla'\cdot\mathbf{M}_{0}\left(\mathbf{r}'\right)\nabla G_{0}\left(\mathbf{r},\mathbf{r}'\right)\right]\mathrm{d}S',\label{eq:SIE_H0}
\end{align}
where the extra terms $\mathbf{E}_{0}^{(i)}$ and $\mathbf{H}_{0}^{(i)}$
account for the incident fields in $\Omega_{0}$, $\mathbf{J}_{0}=\hat{n}_{0}\times\mathbf{H}_{0}$,
$\mathbf{M}_{0}=-\hat{n}_{0}\times\mathbf{E}_{0}$ are the equivalent
electric and magnetic surface current densities which are introduced
mathematically to maintain the tangential fields on $\partial\Omega_{0}$,
and $G_{0}$ is the full-space Green function where the corresponding
wave number is $k_{0}$ with $k_{0}^{2}=\epsilon_{0}\mu_{0}\omega^{2}$.

To summarize, the fields in each region are compactly written as
\begin{align}
\mathbf{E}_{l} & =\delta_{l0}\mathbf{E}_{l}^{(i)}+\mathcal{D}_{l}\mathbf{J}_{l}-\mathcal{K}_{l}\mathbf{M}_{l},\label{eq:SIE_E}\\
\mathbf{H}_{l} & =\delta_{l0}\mathbf{H}_{l}^{(i)}+\mathcal{K}_{l}\mathbf{J}_{l}+\left(\frac{\epsilon_{l}}{\mu_{l}}\right)\mathcal{D}_{l}\mathbf{M}_{l}.\label{eq:SIE_H}
\end{align}
Above, $\delta_{l0}$ is the Kronecker delta and the integro-differential
operators are defined as follows,
\begin{align}
\left\{ \mathcal{D}_{l}\mathbf{f}\left(\mathbf{r}'\right)\right\} \left(\mathbf{r}\right) & =i\omega\mu_{l}\iint_{\partial\Omega_{l}}\left[\mathbf{f}\left(\mathbf{r}'\right)G_{l}\left(\mathbf{r},\mathbf{r}'\right)\right]\mathrm{d}S'\nonumber \\
 & -\frac{1}{i\omega\epsilon_{l}}\iint_{\partial\Omega_{l}}\left[\nabla'\cdot\mathbf{f}\left(\mathbf{r}'\right)\nabla G_{l}\left(\mathbf{r},\mathbf{r}'\right)\right]\mathrm{d}S',\\
\left\{ \mathcal{K}_{l}\mathbf{f}\left(\mathbf{r}'\right)\right\} \left(\mathbf{r}\right) & =-\iint_{\partial\Omega_{l}}\left[\mathbf{f}\left(\mathbf{r}'\right)\times\nabla G_{l}\left(\mathbf{r},\mathbf{r}'\right)\right]\mathrm{d}S',
\end{align}
where $l\in\{0,1\}$ is the domain index. Therefore, for a given point
on the scatterer's surface we have, after taking the operation $\hat{n}_{l}\times$
to Eqs.~(\ref{eq:SIE_E}) and (\ref{eq:SIE_H}), 
\begin{align}
\mathbf{J}_{l} & =\delta_{l0}\mathbf{J}_{l}^{(i)}+\hat{n_{l}}\times\left[\mathcal{K}_{l}\mathbf{J}_{l}+\left(\frac{\epsilon_{l}}{\mu_{l}}\right)\mathcal{D}_{l}\mathbf{M}_{l}\right],\label{eq:SIE_J}\\
\mathbf{M}_{l} & =\delta_{l0}\mathbf{M}_{l}^{(i)}-\hat{n_{l}}\times\left[\mathcal{D}_{l}\mathbf{J}_{l}-\mathcal{K}_{l}\mathbf{M}_{l}\right],\label{eq:SIE_M}
\end{align}
where the current densities $\mathbf{M}_{l}^{(i)}=-\hat{n}_{l}\times\mathbf{E}_{l}^{(i)}$
and $\mathbf{J}_{l}^{(i)}=\hat{n}_{l}\times\mathbf{H}_{l}^{(i)}$
are due to the incident fields in the free space.

The N-M\"uller formulation \citep{makitaloModesResonancesPlasmonic2014,yla-oijalaWellconditionedMullerFormulation2005}
yields stable solutions even at low-frequency limit and it can be
obtained by combining the above two equations with weights $\mu_{l}$
and $\epsilon_{l}$, as well as taking into account the tangential
field continuity at the interface (i.e., $\mathbf{J}=\mathbf{J}_{0}=-\mathbf{J}_{1}$
and $\mathbf{M}=\mathbf{M}_{0}=-\mathbf{M}_{1}$),
\begin{align}
\left(\mu_{0}+\mu_{1}\right)\mathbf{J} & =\left(\mu_{0}\hat{n}_{0}\times\mathcal{K}_{0}+\mu_{1}\hat{n}_{1}\times\mathcal{K}_{1}\right)\mathbf{J}+\mu_{0}\mathbf{J}_{0}^{(i)}\nonumber \\
 & +\left(\epsilon_{0}\hat{n}_{0}\times\mathcal{D}_{0}+\epsilon_{1}\hat{n}_{1}\times\mathcal{D}_{1}\right)\mathbf{M},\\
\left(\epsilon_{0}+\epsilon_{1}\right)\mathbf{M} & =-\left(\epsilon_{0}\hat{n}_{0}\times\mathcal{D}_{0}+\epsilon_{1}\hat{n}_{1}\times\mathcal{D}_{1}\right)\mathbf{J}+\epsilon_{0}\mathbf{M}_{0}^{(i)}\nonumber \\
 & +\left(\mu_{0}\hat{n}_{0}\times\mathcal{K}_{0}+\mu_{1}\hat{n}_{1}\times\mathcal{K}_{1}\right)\mathbf{M}.
\end{align}
This can be further written in the compact form
\begin{equation}
\left(\mathcal{Z}_{0}+\mathcal{Z}_{1}\right)\mathbf{x}=\mathbf{b},\label{eq:SIE_scattering}
\end{equation}
with the vector $\mathbf{x}=[\mathbf{J},\mathbf{M}]$ containing the
equivalent surface current densities to be solved with the knowledge
of the tangential components of the incident fields $\mathbf{b}=[\mu_{0}\mathbf{J}_{0}^{(i)},\epsilon_{0}\mathbf{M}_{0}^{(i)}]$.
We also invoked the matrix notation 
\begin{equation}
\mathcal{Z}_{l}=\begin{bmatrix}\mu_{l} & 0\\
0 & \epsilon_{l}
\end{bmatrix}+\hat{n}_{l}\times\begin{bmatrix}-\mu_{l}\mathcal{K}_{l} & -\epsilon_{l}\mathcal{D}_{l}\\
\epsilon_{l}\mathcal{D}_{l} & -\mu_{l}\mathcal{K}_{l}
\end{bmatrix}.
\end{equation}

\subsection{Eigenmodes}

In the absence of an incident field $\mathbf{b}=0$ holds, and the
system may have nontrivial solutions at certain frequencies (usually
complex-valued), each of them being associated with a resonance mode
of the optical system \citep{makitaloModesResonancesPlasmonic2014,powellResonantDynamicsArbitrarily2014,bernasconiModeAnalysisSecondharmonic2016,lalanneLightInteractionPhotonic2018}.
The resonance frequencies are generally isolated poles in the complex
frequency plane. At a fixed real-valued frequency, however, we can
find the scatterer's eigenmodes \citep{makitaloModesResonancesPlasmonic2014,bernasconiModeAnalysisSecondharmonic2016}
by solving 
\begin{equation}
\left(\mathcal{Z}_{0}+\mathcal{Z}_{1}\right)\mathbf{x}^{(m)}=\lambda^{(m)}\mathbf{x}^{(m)}.\label{eq:SIE_eigenmode_real}
\end{equation}
Rewriting Eq.~(\ref{eq:SIE_eigenmode_real}) in the form of Eq.~(\ref{eq:SIE_scattering}),
we have
\begin{equation}
\left(\mathcal{Z}_{0}+\mathcal{Z}_{1}\right)\frac{\mathbf{x}^{(m)}}{\lambda^{(m)}}=\mathbf{x}^{(m)}.\label{eq:SIE_eigenmode_scattering}
\end{equation}
Comparing Eqs.~(\ref{eq:SIE_scattering}) and (\ref{eq:SIE_eigenmode_scattering}),
it is seen that an incident field $\mathbf{x}^{(m)}$, which is the
$m$th eigenmode, yields a scattered field $\mathbf{x}^{(m)}/\lambda^{(m)}$.
In addition, an eigenmode associated with a smaller eigenvalue $\lambda^{(m)}$
yields a stronger scattered field $\mathbf{x}^{(m)}/\lambda^{(m)}$
when the scatterer is excited by an incident field of $\mathbf{x}^{(m)}$.
Therefore, $1/\lambda^{(m)}$ can be viewed as the \textit{scattering
strength} associated with the $m$th eigenmode.

For a given eigenmode $\mathbf{x}^{(m)}=[\mathbf{J}^{(m)},\mathbf{M}^{(m)}]$,
we can calculate its electric and magnetic fields $\mathbf{E}^{(m)}$
and $\mathbf{H}^{(m)}$ everywhere, by invoking Eqs.~(\ref{eq:SIE_E})
and (\ref{eq:SIE_H}),
\begin{align}
\mathbf{E}_{l}^{(m)}(\mathbf{r}) & =\mathcal{D}_{l}\mathbf{J}_{l}^{(m)}-\mathcal{K}_{l}\mathbf{M}_{l}^{(m)},\label{eq:SIE_Ef}\\
\mathbf{H}_{l}^{(m)}(\mathbf{r}) & =\mathcal{K}_{l}\mathbf{J}_{l}^{(m)}+\left(\frac{\epsilon_{l}}{\mu_{l}}\right)\mathcal{D}_{l}\mathbf{M}_{l}^{(m)},\label{eq:SIE_Hf}
\end{align}
where the subscript $l$ denotes the domain where the field at a given
point $\mathbf{r}\in\Omega_{l}$ is evaluated.

\subsection{Silicon nanodisks}

\subsubsection*{Monomer}

\begin{figure}[tb]
\begin{centering}
\includegraphics{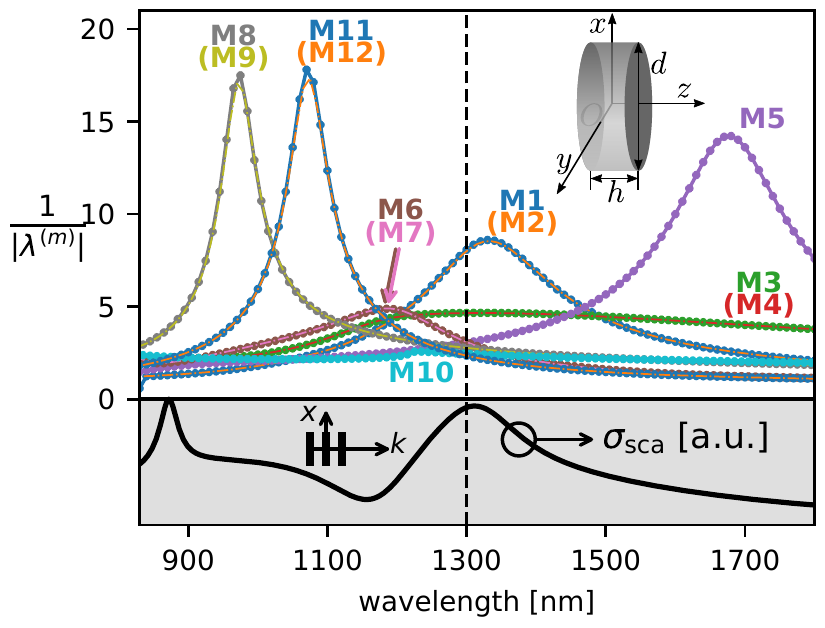}
\par\end{centering}
\caption{The reciprocal of absolute eigenvalues $1/|\lambda^{(m)}|$ of the
first twelve leading eigenmodes in a silicon nanodisk in air as a
function of wavelength. The nanodisk has a thickness of $h=220$ nm
and a diameter of $d=480$ nm and it is situated at the origin of
the Cartesian coordinate system (shown in the inset). Each eigenmode
is labelled by M1 to M12, respectively. In case of degeneracy, one
of the two degenerate eigenmodes is labelled in brackets. These twelve
eigenmodes are ranked by their $1/|\lambda^{(m)}|$ values at wavelength
$1300$ nm (the dashed line). In the filled grey area below the zero
horizontal line, the scattering cross-section ($\sigma_{\mathrm{sca}}$)
in arbitrary units is shown as the black curve when the nanodisk is
illuminated by an $x$-polarized plane wave.}

\centering{}\label{fig:nd-Si-eigval}
\end{figure}
\begin{figure}[tb]
\begin{centering}
\includegraphics[scale=0.15]{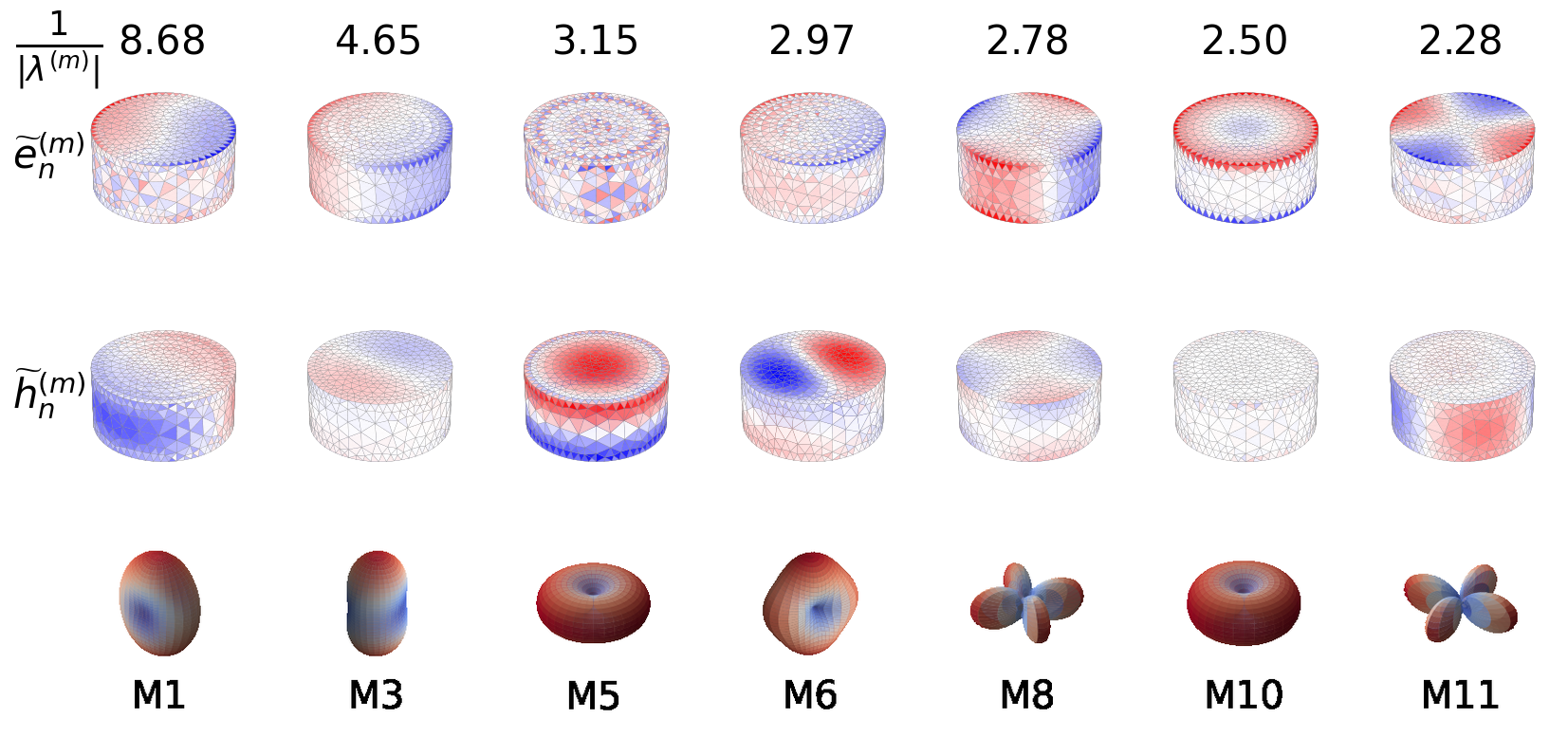}
\par\end{centering}
\begin{centering}
~~~\small{}%
\begin{tabular}{>{\centering}b{31pt}>{\centering}b{31pt}>{\centering}b{31pt}>{\centering}b{31pt}>{\centering}b{31pt}>{\centering}b{31pt}>{\centering}b{31pt}}
\toprule 
$\mathrm{MD}\downarrow$ & $\mathrm{ED}\downarrow$ & $\mathrm{MD}\varodot$ & $\mathrm{MQ}_{\varotimes}^{\varodot}$ & $\mathrm{EQ}_{+-}^{-+}$ & $\mathrm{ED}\varotimes$ & $\mathrm{MQ}_{+-}^{-+}$\tabularnewline
\end{tabular}\normalsize{}
\par\end{centering}
\centering{}\caption{Near fields including the instantaneous electric $\widetilde{e}_{n}^{(m)}$
(first row) and magnetic $\widetilde{h}_{n}^{(m)}$ (second row) fields
on the nanodisk's surface, and far fields (third row) of a few leading
eigenmodes at the wavelength $1300$ nm for the same silicon nanodisk
as in Fig.~\ref{fig:nd-Si-eigval}. Each column is for one eigenmode
with its $1/|\lambda^{(m)}|$ value given on the top.}
\label{fig:nd-Si-nfpl}
\end{figure}
\begin{figure*}[t]
\centering{}\includegraphics[scale=0.18]{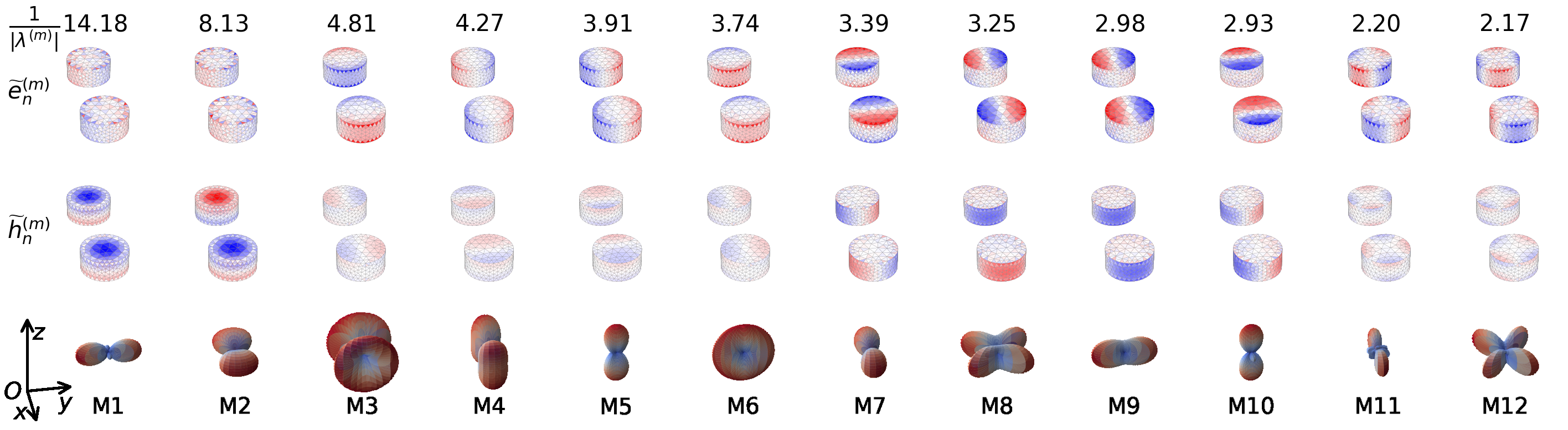}\caption{Near and far fields of eigenmodes M1--12 at 1600 nm in a dimer consisting
of two identical silicon nanodisks (same as in Fig.~\ref{fig:nd-Si-eigval})
with a center-to-center gap of $480$ nm.}
\label{fig:nd-dimer-Si-nfpl}
\end{figure*}
\begin{figure*}
\begin{centering}
\includegraphics[scale=0.18]{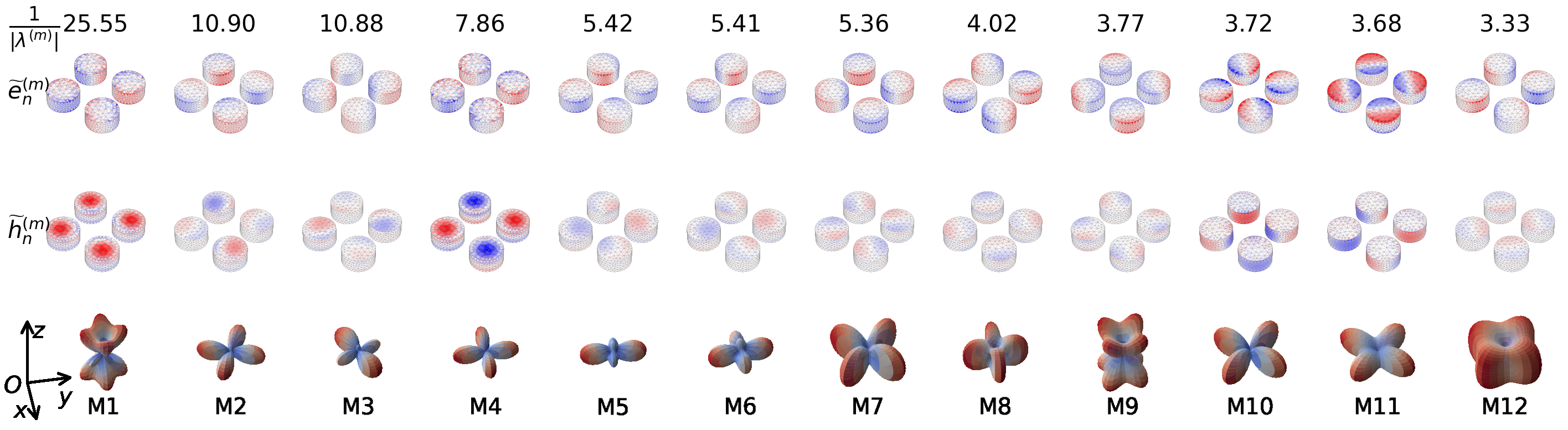}
\par\end{centering}
\centering{}\caption{Near and far fields of a few leading eigenmodes M1--12 at 1600 nm
in a tetramer consisting of four identical silicon nanodisks (same
as in Fig.~\ref{fig:nd-Si-eigval}) with the center-to-center gap
of two opposite neighbouring nanodisks being $480$ nm.}
\label{fig:nd-tetramer-Si-nfpl}
\end{figure*}
For the purpose of demonstrating the eigenmodes in all-dielectric
nanostructures, we study a silicon nanodisk in air. The nanodisk has
a thickness of $h=220$ nm and a diameter of $d=480$ nm and it is
situated at the origin of the Cartesian coordinate system as shown
in the inset of Fig.~\ref{fig:nd-Si-eigval}. The eigenvalues are
solved in the wavelength range of $[850,1750]$ nm and the twelve
dominant eigenmodes are shown in Fig.~\ref{fig:nd-Si-eigval}. The
eigenmodes sharing the same $1/|\lambda^{(m)}|$ values are degenerate
modes, whose fields are of identical distribution but distinctively
oriented. Treating the degenerate eigenmodes together, we observe
seven unique eigenmodes showing different resonance responses. Mode
M1 (M2) has a resonance wavelength around 1330 nm. Mode M3 (M4) does
not show clear resonance as its scattering strength value is flattened
over a relatively wide spectral range. Mode M5 is a non-degenerate
mode showing resonance scattering at 1675 nm. Mode M6 (M7) has a relatively
weak resonance near 1200 nm. Resonances at two shorter wavelengths
970 nm and 1070 nm are associated with the modes M8 (M9) and M11 (M12),
respectively. At last, mode M10 is weakly resonant across the entire
spectral range of interest. At this point, we emphasize that the resonance
wavelengths determined from the scattering strength values are the
genuine resonance wavelengths that do not depend on any external incident
field. The calculated scattering cross section in arbitrary units
from the nanodisk when illuminated by an $x$-polarized plane wave
propagating along the $z$-axis is shown as the black curve in the
filled grey area. The scattering cross section under the illumination
of a linearly polarized plane wave shows a resonance peak near 1300
nm (the dashed line), at which we have ranked the aforementioned eigenmodes
by their scattering strength values $1/|\lambda^{(m)}|$.

The corresponding near and far fields of the eigenmodes at the wavelength
of $1300$ nm are inspected in Fig.~\ref{fig:nd-Si-nfpl}, and only
one of degenerate eigenmodes, if any, is considered. For visualizing
both the electric and magnetic near fields with balanced amplitudes
\citep{coltonIntegralEquationMethods2013,taskinenImplementationFormulationElectromagnetic2006},
we introduce the scaled instantaneous electric and magnetic fields
as
\begin{equation}
\widetilde{\mathbf{e}}^{(m)}\left(\mathbf{r},t\right)=\sqrt{\epsilon_{0}}\mathrm{Re}\left\{ \mathbf{E}_{0}^{(m)}(\mathbf{r})e^{-i\omega t}\right\} \label{eq:efield-scaled}
\end{equation}
and
\begin{equation}
\widetilde{\mathbf{h}}^{(m)}\left(\mathbf{r},t\right)=\sqrt{\mu_{0}}\mathrm{Re}\left\{ \mathbf{H}_{0}^{(m)}(\mathbf{r})e^{-i\omega t}\right\} ,\label{eq:hfield-scaled}
\end{equation}
respectively, where $t$ is a time instant, $r\in\partial\Omega_{0}$,
and $\mathrm{Re}\{\cdot\}$ denotes the real part. It is noted that
the instantaneous electric and magnetic energy densities will be written
as halves the square of the scaled instantaneous electric and magnetic
fields, i.e., $w_{e}=[\widetilde{\mathbf{e}}^{(m)}]^{2}/2$ and $w_{h}=[\widetilde{\mathbf{h}}^{(m)}]^{2}/2$,
respectively \citep{jacksonClassicalElectrodynamics1998}. On nanodisk's
surface, the normal component of the scaled field $\widetilde{e}_{n}^{(m)}=\hat{n}\cdot\widetilde{\mathbf{e}}^{(m)}$
at a time instant $t_{0}$ is shown for the eigenmode's electric field,
and the magnetic counterpart is shown by $\widetilde{h}_{n}^{(m)}=\hat{n}\cdot\widetilde{\mathbf{h}}^{(m)}$
at a $\pi/2$-phase-delayed time instant $t_{0}'$ with $\omega(t_{0}'-t_{0})=\pi/2$,
at which time the magnetic field generally reaches its maximum.

For a nanodisk monomer, the surface normal near fields $\widetilde{e}_{n}^{(m)}$
and $\widetilde{h}_{n}^{(m)}$ are shown in the first and second rows,
respectively, in Fig.~\ref{fig:nd-Si-nfpl}, and the eigenmodes'
corresponding far fields are shown in the third row. Each column displays
the near and far fields for one eigenmode with its $1/|\lambda^{(m)}|$
value given on the top of each column. As can be seen from the radiation
patterns, the eigenmode M1 is a dipolar mode showing a typical doughnut
shape. Furthermore, the near fields show that the mode M1 is of magnetic
dipole with a strong transverse magnetic field, which is denoted by
the symbol MD$\downarrow$ with the downarrow representing a field
vector in the transverse plane. In the second column, M3 has a smaller
$1/|\lambda^{(3)}|$ value, and the radiation pattern and near fields
indicate that it is an electric dipolar mode, which is denoted by
ED$\downarrow$. Under a similar observation, it turns out that M5
is a magnetic dipolar mode that differs from M1 by a dominating $z$
magnetic field component, denoted by MD$\varodot$ with the symbol
``$\varodot$'' implying a field vector along the positive $z$
axis. Mode M6 emerges from a magnetic quadrupole lying in a plane
parallel to the $z$ axis, which is symbolized by MQ$_{\varotimes}^{\varodot}$
where the symbol ``$\varotimes$'' implies a field vector along
the negative $z$ axis that is $\pi$-out-of-phase with respect to
the ``$\varodot$'' field vector. The mode M6 is distinct from another
magnetic quadrupolar mode M11 where the quadrupole resides in the
$xy$ plane, denoted by MQ$_{+-}^{-+}$ where the symbols ``$-$''
and ``$+$'' represent negative and positive field extremes, respectively.
An electric quadrupolar mode, denoted by EQ$_{+-}^{-+}$, is observed
in mode M8. At last, mode M10 is an electric dipolar mode, denoted
by ED$\varotimes$, with a dominating electric $z$ field component.

\subsubsection*{Oligomers}

In silicon nanodisks ensembles, i.e., nanodisk oligomers consisting
of several nanodisks, further local field enhancements and mode hybridization
happen via optical near-field coupling. Here, we also investigate
the near fields of a few leading eigenmodes in nanostructure assemblies
that consist of two and four identical nanodisks, so-called nanodisk
dimer and tetramer, respectively.

The eigenmodes' near and far fields of a dimer are shown in Fig.~\ref{fig:nd-dimer-Si-nfpl}.
Modes M1 and M2 are hybridized from two longitudinal magnetic dipolar
modes, where the former arises from two parallel longitudinal MD modes,
and the latter is due to two anti-parallel longitudinal MD modes.
Transversal magnetic modes include M7-10, that are due to the hybridization
of, subsequently, two $y$-oriented anti-parallel transversal MD modes,
two $x$-oriented anti-parallel transversal MD modes, two $x$-oriented
parallel transversal MD modes, and two $y$-oriented parallel transversal
MD modes. The electric counterparts of the magnetic modes M7--10
are observed in transversal electric modes M3--6. In detail, M3 is
hybridized from two $x$-oriented anti-parallel transversal ED modes,
M4 is due to two $y$-oriented anti-parallel transversal ED modes,
M5 arises from two $y$-oriented parallel transversal ED modes, and
two $x$-oriented parallel transversal ED modes yields the hybrid
mode M6. Additionally, two transveral EQ modes, when arranged differently,
hybridize into two distinct modes M11 and M12.

In a tetramer, modes hybridization leads to more types of eigenmodes
as a result of more near-field coupling channels. Near and far fields
of twelve eigenmodes with leading scattering strengths $1/|\lambda^{(m)}|$
are shown in Fig.~\ref{fig:nd-tetramer-Si-nfpl}. It is clear that
mode M1 arises from four parallel longitudinal magnetic modes, whereas
the longitudinal magnetic fields in every two neighbouring nanodisks
in M4 are $\pi$-out-of-phase with respect to each other. Modes hybridized
from four transversal ED modes contain M7 where the ED moments are
$\pi$-out-of-phase along the radial direction between every two neighbouring
elements, M8 where the ED moments are $\pi$-out-of-phase between
every two neighbouring elements along the azimuthal direction, M9
with radially in-phase ED moment in each element, and M12 with in-phase
ED moment in each nanodisk along the azimuthal direction. Hybridization
of transversal MDs leads to mode M10 with the MD moments being along
the radial direction and $\pi$-out-of-phase between the neighbouring
nanodisks and M11 with azimuthally in-phase MD moments. In addition,
M10 and M11 are the magnetic counterparts of modes M7 and M12. It
is not straightforward to gain physical insights of the origins of
hybridized modes M2, M3, M5, and M6 without future investigation,
and determining the near-field coupling channels for these modes is
less important and not further discussed in this work.

As it can be seen from Figs.~\ref{fig:nd-Si-nfpl}, \ref{fig:nd-dimer-Si-nfpl},
and \ref{fig:nd-tetramer-Si-nfpl}, an eigenmode is associated with
its unique near-field distribution and far-field radiation pattern.
Tuning light scattered from nano-objects via selective excitation(s)
of certain eigenmode(s) could serve as a basic building block for
engineering optical near and/or far field(s). It is, therefore, vital
to design an excitation field which highly overlaps with the field
of a specific eigenmode for an effective mode excitation, i.e., mode-matching
field.

\section{Inverse design via tight focusing\label{sec:inverse-focusing}}

As shown in previous Sec.~\ref{sec:eigenmodes_formulation}, the
eigenmode's near fields in optical nanoantennas are genuinely localized
and vectorial at the nanoscale. A potential free-space approach to
design such a 3D mode-matching vector field is to shape a focal field
(that matches with the eigenmode's near field in optical nanoantennas
that are to be placed in the vicinity of the focal point), by focusing
a paraxial vector beam in an aplanatic system with high NA. In the
previous work, tightly focused cylindrical vector beams of fundamental
order, i.e., radially and azimuthally polarized beams \citep{gomezDarkSidePlasmonics2013,sancho-parramonDarkModesFano2012},
and of higher-order polarization states \citep{zang*EfficientHybridmodeExcitation2021}
have been used for optical excitation of dark and hybrid modes in
optical nanoantennas. In these forward design approaches, the focused
field has a predefined field profile and only qualitative match of
polarization distribution between the hybrid modes and the excitation
fields was considered. To design the best possible focal field that
quantitatively matches with the eigenmode's field, it is necessary
to solve an inverse problem where the needed paraxial vector beam
to form a desired focal field after tight focusing is to be sought.

\subsection{Tight focusing: forward and backward propagation}

\begin{figure}[tb]
\centering{}\includegraphics[scale=0.72]{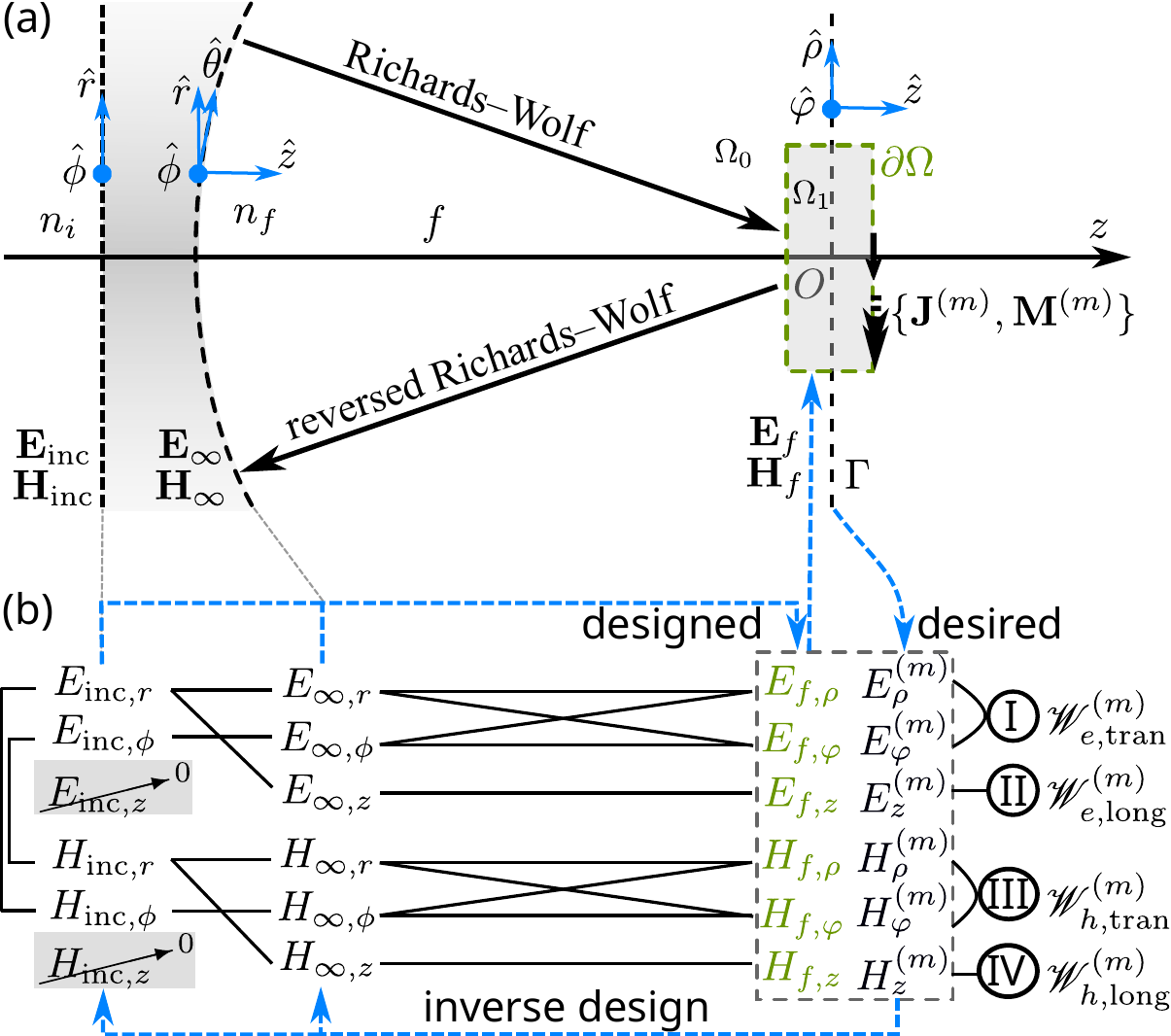}\caption{The inverse design scheme of focused vector beams with effective mode-matching
profiles. (a) The forward and backward tight focusing processes. The
nanostructure's surface is denoted by $\partial\Omega$, where the
$m$th eigenmode's equivalent surface current density $\{\mathbf{J}^{(m)},\mathbf{M}^{(m)}\}$
resides. The middle symmetry plane $\Gamma$ coincides with the focal
plane ($z=0$ plane), where the modal field $\{\mathbf{E}^{(m)},\mathbf{H}^{(m)}\}$
is evaluated. The reference field $\mathbf{E}_{\infty}$ resides on
the reference spherical cap with a radius being equal to the focal
length $f$. The incident beam is denoted by $\mathbf{E}_{\mathrm{inc}}$.
The nanostructure occupies domain $\Omega_{1}$, whereas the free
space is denoted by $\Omega_{0}$ with a refractive index of $n_{f}$.
The index before the reference sphere is denoted by $n_{i}$. (b)
The inverse design flowchart. Solid black lines indicate connections
between the associated field components. The modal field $\{\mathbf{E}^{(m)},\mathbf{H}^{(m)}\}$
on the focal plane $\Gamma$ serves as the desired focal field, from
which the the pupil field $\{\mathbf{E}_{\mathrm{inc}},\mathbf{H}_{\mathrm{inc}}\}$
and thus the focal field $\{\mathbf{E}_{f},\mathbf{H}_{f}\}$ are
designed. Desired focal fields are categorized into four groups by
the time-averaged energies stored in the transverse and longitudinal,
electric and magnetic fields in the focal plane, denoted by $\ensuremath{\mathscr{W}_{e,\mathrm{tran}}^{(m)}}$,
$\ensuremath{\mathscr{W}_{e,\mathrm{long}}^{(m)}}$, $\ensuremath{\mathscr{W}_{h,\mathrm{tran}}^{(m)}}$,
and $\ensuremath{\mathscr{W}_{h,\mathrm{long}}^{(m)}}$, respectively.}
\label{fig:inverse}
\end{figure}

Before proceeding, we first review the forward and backward propagation
in tight focusing of vector beams. In detail, we choose a coordinate
system such that the middle plane of the nanostructure(s) coincides
with the focal plane $\Gamma$ (where $z=0$), as shown in Fig.~\ref{fig:inverse}.
For a given eigenmode, its near field $\{\mathbf{E}^{(m)},\mathbf{H}^{(m)}\}$
is evaluated from Eqs.~(\ref{eq:SIE_Ef}) and (\ref{eq:SIE_Hf})
in the focal plane $\Gamma$. For the eigenmode's electric near field
to be mode-matched, a 3D vector field $\mathbf{E}_{f}$ is designed
via the tight focusing of a paraxial vector beam $\mathbf{E}_{\mathrm{inc}}$
at the pupil plane (which is mapped to $\mathbf{E}_{\infty}$ on the
reference sphere).

The tight focusing process is governed by the Richards--Wolf formalism
\citep{wolfElectromagneticDiffractionOptical1959,richardsElectromagneticDiffractionOptical1959,pereiraSuperresolutionMeansPolarisation2004},
and the focal field at a point $\mathbf{R}_{f}=(\rho,\varphi,z)$
in cylindrical coordinates writes as
\begin{align}
\mathbf{E}_{f}(\rho,\varphi,z) & =\frac{-ik}{2\pi}\int\limits _{0}^{\theta_{\mathit{\mathrm{max}}}}\int\limits _{0}^{2\pi}\mathbf{E}_{\infty}(\theta,\phi)\nonumber \\
 & \quad\times e^{i\mathbf{k}\cdot\mathbf{R}_{f}}\sin\theta\mathrm{d}\phi\mathrm{d}\theta,\label{eq:efocal}
\end{align}
where the wave vector $\mathbf{k}$ is along the direction from a
point on the reference sphere $(\phi,\theta)$ to the focal point
and the corresponding wave number $k=k_{0}$ (i.e., the same value
as in domain $\Omega_{0}$). The maximum converging angle $\theta_{\mathrm{max}}=\arcsin(\mathrm{NA}/n_{f})$
is limited by the $\mathrm{NA}$ of the aplanatic system, where $n_{f}$
is the refractive index after the reference sphere. We can rewrite
Eq.~(\ref{eq:efocal}) into Fourier transform in Cartesian coordinates
\citep{Leutenegger200611277,zhangInverseMethodEngineer2018} 
\begin{align}
\mathbf{E}_{f}(x,y,z) & =\frac{-ik}{2\pi}\int\limits _{-\infty}^{+\infty}\int\limits _{-\infty}^{+\infty}\left[\frac{\mathbf{E}_{\infty}(k_{x},k_{y})e^{ikz\cos\theta}}{k^{2}\cos\theta}\right]\nonumber \\
 & \quad\times e^{-ik_{x}x}e^{-ik_{y}y}\mathrm{d}k_{x}\mathrm{d}k_{y},\label{eq:efocal_FT}
\end{align}
where $k_{x}=-k\sin\theta\cos\phi$ and $k_{y}=-k\sin\theta\sin\phi$
are the $x$ and $y$ components of the wave number, respectively.
Here, the surface integral runs from $-\infty$ to $+\infty$, as
the field $\mathbf{E}_{\infty}$ on the reference sphere is zero outside
of the NA. In order to obtain the necessary paraxial vector beam or
the field $\mathbf{E}_{\infty}$ on the reference sphere to generate
the desired 3D mode-matching vector field $\mathbf{E}_{f}$, we perform
the inverse Fourier transform for the above Eq.~(\ref{eq:efocal_FT})
except $\mathbf{E}_{f}$ is replaced by the eigenmode's near field
in the focal plane $\mathbf{E}^{(m)}\vert_{\Gamma}$,
\begin{align}
\frac{\mathbf{E}_{\infty}\left(k_{x},k_{y}\right)e^{ikz\cos\theta}}{\cos\theta} & =\frac{ik}{2\pi}\int\limits _{-\infty}^{+\infty}\int\limits _{-\infty}^{+\infty}\left.\mathbf{E}^{(m)}\right|_{\Gamma}\nonumber \\
 & \quad\times e^{ik_{x}x}e^{\mathrm{i}k_{y}y}\mathrm{d}x\mathrm{d}y,\label{eq:eref_iFT}
\end{align}
which can be considered as the governing equation for a reversed Richards--Wolf
tight focusing process.

Ideally, the paraxial input vector beam is inversely restored from
the reference field \citep{wolfScalarRepresentationElectromagnetic1959,zang*EfficientHybridmodeExcitation2021}

\begin{align}
\mathbf{E}_{\mathrm{inc}}(r,\phi) & =\mathbf{L}^{\mathrm{-1}}(\theta)\mathbf{E}_{\infty}(\theta,\phi),\label{A-eq:einc}
\end{align}
where the radius is $r=f\sin\theta$ with $f$ being the focal length
and the mapping matrix in cylindrical coordinates is \citep{zang*EfficientHybridmodeExcitation2021}
\begin{equation}
\mathbf{L}^{\mathrm{-1}}(\theta)=\frac{1}{f\sqrt{\cos\theta}}\sqrt{\frac{n_{f}}{n_{i}}}\begin{bmatrix}\cos\theta & 0 & \sin\theta\\
0 & 1 & 0\\
-\sin\theta & 0 & \cos\theta
\end{bmatrix},
\end{equation}
where $n_{i}$ is the refractive index before the reference sphere.
Above, we followed Richards and Wolf \citep{richardsElectromagneticDiffractionOptical1959}
for the refraction at the aplanatic lens, the reference field amplitude
vector writes as $\mathbf{E}_{\infty}(\theta,\phi)=f\sqrt{n_{i}/n_{f}}\sqrt{\cos\theta}[E_{\mathrm{inc},r}\hat{\theta}+E_{\mathrm{inc},\phi}\hat{\phi}+E_{\mathrm{inc},z}(\hat{\theta}\times\hat{\phi})]$.

\subsection{Degrees of freedom\label{subsec:dof}}

The associated magnetic fields are governed by similar formulas as
in Eqs.~(\ref{eq:efocal})--(\ref{A-eq:einc}) for the electric
counterparts, but for brevity we do not repeat them here. However,
it is noted that in Richards--Wolf formalism the magnetic fields
are linked to electric fields by revoking the relations $\mathbf{H}_{\mathrm{inc}}=1/Z_{i}(\mathbf{k}_{i}/k_{i})\times\mathbf{E}_{\mathrm{inc}}$
and thus $\mathbf{H}_{\infty}=1/Z_{f}(\mathbf{k}_{f}/k_{f})\times\mathbf{E}_{\infty}$
\citep{wolfElectromagneticDiffractionOptical1959} as a result of
beam-like characteristic of the vector fields on the pupil plane and
on the reference sphere. Here, $Z_{i}$ and $Z_{f}$ are the associated
wave impedances. This implies that only two components in the pupil
vector field are completely independent. In cylindrical coordinates,
four possible combinations of two field components are $(E_{\mathrm{inc},r},E_{\mathrm{inc},\phi})$,
$(H_{\mathrm{inc},r},H_{\mathrm{inc},\phi})$, $(E_{\mathrm{inc},r},H_{\mathrm{inc},r})$,
and $(E_{\mathrm{inc},\phi},H_{\mathrm{inc},\phi})$, and choosing
any pair will fully determine the vector field on the pupil plane,
then on the reference sphere, and finally in the focal region.

The fact that the incident vector beam is paraxial ($E_{\mathrm{inc},z}\approx0$
and $H_{\mathrm{inc},z}\approx0$) imposes a restriction on the radial
and longitudinal components of the reference electric \citep{zhangInverseMethodEngineer2018}
and magnetic fields via the following relations:
\begin{align}
-\sin\theta E_{\infty,r}(\theta,\phi)+\cos\theta E_{\infty,z}(\theta,\phi) & =0,\label{eq:restriction-E}\\
-\sin\theta H_{\infty,r}(\theta,\phi)+\cos\theta H_{\infty,z}(\theta,\phi) & =0.\label{eq:restriction-H}
\end{align}
As a consequence, the focused electric field $\mathbf{E}_{f}$ bears
all three components that are nevertheless not completely independent
\citep{foremanInversionDebyeWolfDiffraction2008}. In other words,
all three components of the electric focal field are determined uniquely
by two components of the beam-like pupil field $\mathbf{E}_{\mathrm{inc}}$,
as seen in Fig.~\ref{fig:inverse}(b) where the black solid lines
indicate the dependencies of field components on different planes.
In order to inversely design a focal field that is truly generated
by a beam-like vector pupil field, inherent dependencies of different
field components of the desired focal field should be carefully investigated.

\subsection{Inverse design strategies}

To inversely design a focal field that potentially matches with an
eigenmode's field, i.e., the desired focal field, in principle all
the electric and magnetic modal field components should be taken into
account. However, bearing in mind the degrees of freedom discussed
in the previous section, an inconsistent pupil field could result
from a backward propagation that ignores the inherent dependencies
of all components \citep{foremanInversionDebyeWolfDiffraction2008}
in the desired focal field. In contrast, one has to choose appropriate
components in the desired focal field that will unambiguously determine
the beam-like pupil field and thus the designed focal field. It is
observable in Figs.~\ref{fig:nd-Si-nfpl}, \ref{fig:nd-dimer-Si-nfpl},
and \ref{fig:nd-tetramer-Si-nfpl} that only some of the electric
and magnetic field components are dominant in an eigenmode's near
field. For instance, the mode M10 of a single nanodisk in Fig.~\ref{fig:nd-Si-nfpl}
shows strong longitudinal electric field component. Moreover, in Fig.~\ref{fig:nd-dimer-Si-nfpl}
modes M3--6 of a nanodisk dimer are clearly dominated by their transverse
electric and longitudinal magnetic field components, whereas the transverse
magnetic and longitudinal electric field components are dominant in
modes M7--10 of a nanodisk dimer.

For a more quantitative analysis, we categorize the desired focal
fields according to the time-averaged energies associated with the
transverse and longitudinal electric and magnetic fields in the focal
plane, i.e., the following surface integrals of the energy densities
\citep{snyderOpticalWaveguideTheory1983},
\begin{align}
\mathscr{W}_{e,\mathrm{tran}}^{(m)} & =\frac{1}{2}\epsilon_{0}\int_{\Gamma}\left(\left|E_{\rho}^{(m)}\right|^{2}+\left|E_{\varphi}^{(m)}\right|^{2}\right)\mathrm{d}S,\\
\mathscr{W}_{e,\mathrm{long}}^{(m)} & =\frac{1}{2}\epsilon_{0}\int_{\Gamma}\left|E_{z}^{(m)}\right|^{2}\mathrm{d}S,\\
\mathscr{W}_{h,\mathrm{tran}}^{(m)} & =\frac{1}{2}\mu_{0}\int_{\Gamma}\left(\left|H_{\rho}^{(m)}\right|^{2}+\left|H_{\varphi}^{(m)}\right|^{2}\right)\mathrm{d}S,\\
\mathscr{W}_{h,\mathrm{long}}^{(m)} & =\frac{1}{2}\mu_{0}\int_{\Gamma}\left|H_{z}^{(m)}\right|^{2}\mathrm{d}S.
\end{align}
In principle, the surface integration is performed over an infinite
focal plane $\Gamma$, but in numerical evaluation we have to truncate
the integration surface in an area where the desired focal field remains
significant and discard the field elsewhere.

In group \textcircled{\tiny I}, the focal-plane energy associated
with the transverse electric field is larger than those of the longitudinal
electric field component as well as both focal-plane energies related
to the transverse and longitudinal magnetic field components. In group
\textcircled{\tiny II}, the focal-plane energy of the longitudinal
electric field overwhelms the other three types of energies. The magnetic
counterparts of groups \textcircled{\tiny I} and \textcircled{\tiny II}
are group \textcircled{\tiny III} and group \textcircled{\tiny IV},
respectively. Concisely, we may write
\begin{align}
 & \mathrm{max}(\mathscr{W}_{e,\mathrm{tran}}^{(m)},\mathscr{W}_{e,\mathrm{long}}^{(m)},\mathscr{W}_{h,\mathrm{tran}}^{(m)},\mathscr{W}_{h,\mathrm{long}}^{(m)})\nonumber \\
= & \begin{cases}
\mathscr{W}_{e,\mathrm{tran}}^{(m)}, & \text{in group \textcircled{\tiny I}},\\
\mathscr{W}_{e,\mathrm{long}}^{(m)}, & \text{in group \textcircled{\tiny II}},\\
\mathscr{W}_{h,\mathrm{tran}}^{(m)}, & \text{in group \textcircled{\tiny III}},\\
\mathscr{W}_{h,\mathrm{long}}^{(m)}, & \text{in group \textcircled{\tiny IV}}.
\end{cases}
\end{align}
For eigenmodes belonging to the group \textcircled{\tiny I}, the
corresponding transverse electric field is chosen as the candidate
component of the desired focal field, upon which the radial and azimuthal
components of the reference field $(E_{\infty,r},E_{\infty,\phi})$
and thus the pupil field $(E_{\mathrm{inc},r},E_{\mathrm{inc},\phi})$
are determined. The longitudinal electric focal field is, however,
a free parameter so that the longitudinal electric field component
on the reference sphere $E_{\infty,z}$ is determined from the radial
component $E_{\infty,r}$ according to the restriction in Eq.~(\ref{eq:restriction-E}).
In this regard, the pupil vector field is guaranteed to be beam-like,
and the designed focal field including the magnetic part is fully
determined by the pair of field components $(E_{\mathrm{inc},r},E_{\mathrm{inc},\phi})$.

For eigenmodes in the group \textcircled{\tiny II}, the longitudinal
electric field component of the desired focal field $E_{z}^{(m)}$
determines the longitudinal electric reference field $E_{\infty,z}$,
then the radial electric pupil field $E_{\mathrm{inc},r}$ using the
restriction in Eq.~(\ref{eq:restriction-E}) where the radial electric
reference field is a free parameter. What remains to be determined
is either the azimuthal electric or the radial magnetic pupil field.
But, the azimuthal electric pupil field is only determined by the
azimuthal electric reference field, which in turn depends on the radial
and azimuthal desired electric focal field both of which are usually
trivial for an eigenmode showing strong longitudinal electric field.
On the other hand, a strong longitudinal electric field component
of the desired focal field is usually accompanied by slightly less
stronger transverse magnetic field component $H_{\rho}^{(m)}$ and/or
$H_{\varphi}^{(m)}$ from which the radial magnetic reference field
$H_{\infty,r}$ and then the radial magnetic pupil field $H_{\mathrm{inc},r}$
can be determined. The magnetic counterparts of the aforementioned
electric design strategies apply to the groups \textcircled{\tiny III}
and \textcircled{\tiny IV}. At this point, it is worth pointing out
that the restriction on the field components $E_{\infty,r}$ and $E_{\infty,z}$,
as well as $H_{\infty,r}$ and $H_{\infty,z}$ poses a challenge to
design a 3D focused vector beam, whose electric and magnetic field
components perfectly match the eigenmode's near field.

Besides the constraint imposed by the beam-like characteristics in
the pupil field, another limitation in the inverse design of a focal
field is due to the fact that the reference field is band-limited,
i.e., spatial frequencies that are larger than the free-space wave
number are evanescent and cannot be backward propagated onto the reference
sphere. This is associated with the fact that the tight focusing process
relies on free-space beam propagation, and the focal field is actually
the far field propagated from the secondary field on the reference
sphere. In addition, back propagated waves are further limited by
the NA of the aplanatic system.

\begin{figure}[t]
\centering{}\includegraphics[scale=0.72]{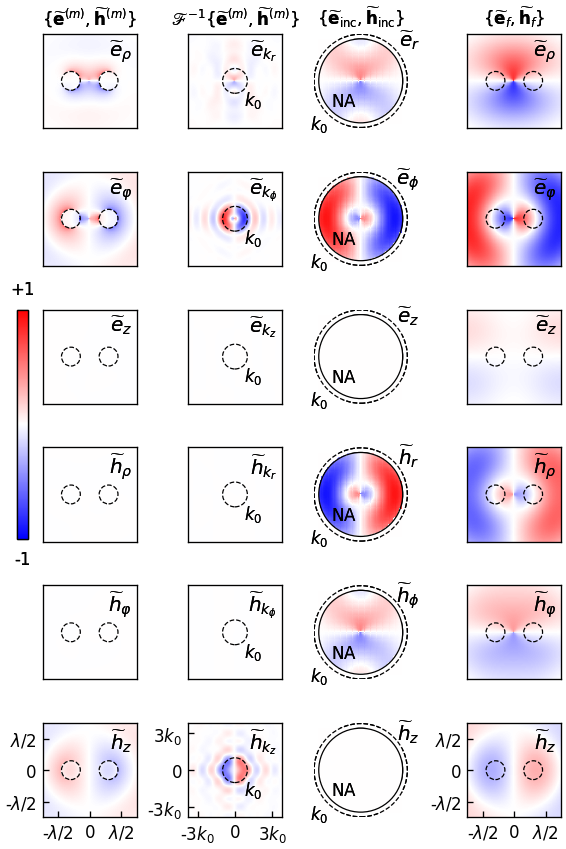}\caption{Scaled electric and magnetic field components {[}defined in Eqs.~(\ref{eq:efield-scaled})
and (\ref{eq:hfield-scaled}){]} at different stages in the inverse
design for M2 in the nanodisk dimer whose eigenmodes are shown in
Fig.~\ref{fig:nd-dimer-Si-nfpl}. The dashed circles represent the
nanodisk's outlines in the desired (first column) and designed (fourth
column) focal fields, or the threshold of maximum free-space wave
number $k_{0}$ in $\mathbf{k}$ space of the desired focal field
(second column) and the designed pupil field (third column). The solid
circle in the designed pupil field indicates the wave number limited
by NA with a value of 0.9 used in this inverse design.}
\label{fig:nd-dimer-pupil_inv-ieig2}
\end{figure}

An example eigenmode belonging to group \textcircled{\tiny I} is
the mode M2 in the nanodisk dimer, and its scaled electric and magnetic
fields at different stages of the inverse design are shown in Fig.~\ref{fig:nd-dimer-pupil_inv-ieig2}.
The first and second columns show all six field components of the
desired focal field in the real space and spatial frequency $\mathbf{k}$
space, respectively. The inversely designed pupil field's components
are shown in the third column, and the designed focal field's components
can be seen in the fourth column. Both the radial and azimuthal scaled
electric field components of the designed focal field effectively
match with the desired eigenmode's near field, even though the desired
focal field is so locally confined that a considerable amount of light
is at high spatial frequencies outside the $k_{0}$-circle and has
to be discarded in the free-space backward propagation of the inverse
design. Imposed by the beam-like characteristics of the designed pupil
field, the azimuthal scaled magnetic field is identical to the radial
scaled electric field, whereas the radial scaled magnetic field is
$\pi$-phase shifted with respect to the azimuthal scaled electric
field. Four other field components are also present as a result of
tight focusing of the beam-like pupil field and they are dependent
on the radial and azimuthal electric field components {[}see Fig.~\ref{fig:inverse}(b)
for the connections of field components{]}. In particular, the designed
longitudinal magnetic field is $\pi$-out-of-phase with respect to
the desired longitudinal magnetic field. Nevertheless, as it will
be shown later with the expansion coefficients in the next section,
this designed focal field still effectively excites the mode M2 in
the nanodisk dimer.

\begin{figure}[t]
\centering{}\includegraphics[scale=0.72]{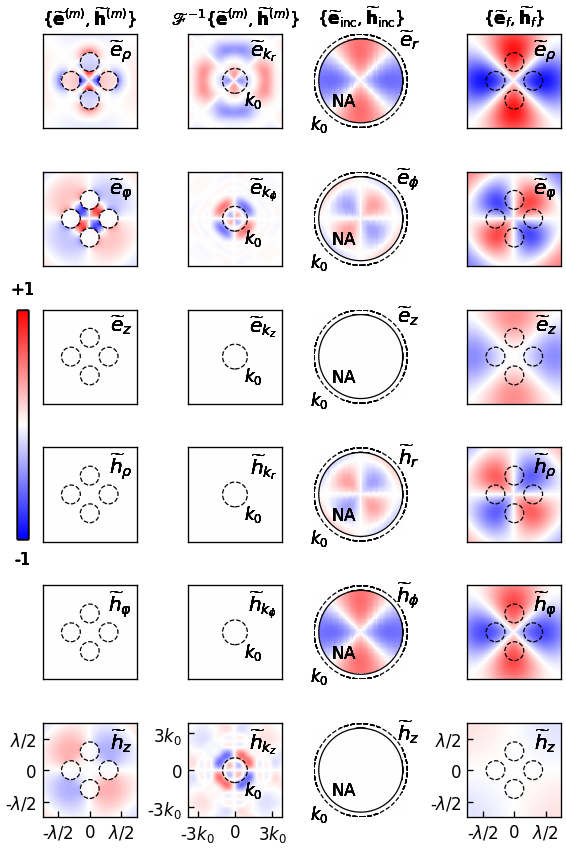}\caption{Scaled electric and magnetic field components in the inverse design
for M7 in the nanodisk tetramer (similar to Fig.~\ref{fig:nd-dimer-pupil_inv-ieig2})
whose eigenmodes are shown in Fig.~\ref{fig:nd-tetramer-Si-nfpl}.}
\label{fig:nd-tetramer-pupil_inv-ieig7}
\end{figure}

Another example is shown in Fig.~\ref{fig:nd-tetramer-pupil_inv-ieig7}
for inversely designing a focal field to match with the eigenmode
M7 in the nanodisk tetramer. An effective match between the designed
and the desired focal fields is achieved in radial and azimuthal electric
field components, as well as the longitudinal magnetic field component.
Again, as it will be shown later with the expansion coefficients in
the next section, this designed focal field can effectively and exclusively
excite the mode M7 in the nanodisk tetramer. In this case, it seems
dropping waves with spatial frequencies higher than those limited
by NA has a trivial effect on the excitation efficiency. A full list
of scaled field components in the inverse design of focal field for
all twelve eigenmodes in the nanodisk monomer, dimer, and tetramer
is compiled in the Supplementary Material (SM).

\section{Mode expansion coefficients\label{sec:excitation_coeff}}

\begin{figure*}[t]
\begin{centering}
\includegraphics{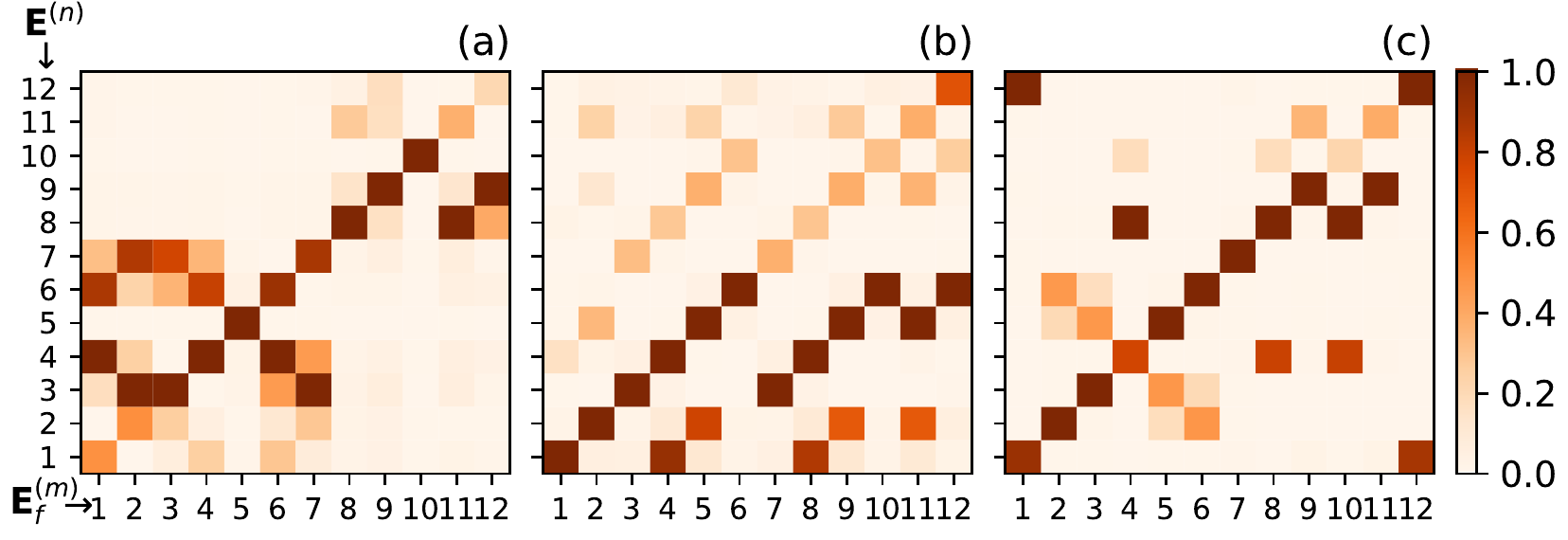}
\par\end{centering}
\caption{The mode excitation coefficients between the designed focal fields
$\mathbf{E}_{f}^{(m)}$ (along the horizontal axis) and eigenmodes
$\mathbf{E}^{(n)}$ (along the vertical axis) for the nanodisk (a)
monomer, (b) dimer, and (c) tetramer. For each designed focal field,
the expansion coefficients for the first twelve eigenmodes are normalized
to the maximum value.}

\centering{}\label{fig:exp_coeffs}
\end{figure*}
Following the analysis of eigenmodes in nano-objects (Sec.~\ref{sec:eigenmodes_formulation})
and the inverse design of tightly focused beam (Sec.~\ref{sec:inverse-focusing}),
it is important to see how well a designed focal field matches with
an eigenmode's near field. A figure of merit is the modal expansion
coefficient which characterizes the overlapping between the inversely
designed vector beam and the eigenmode's near field. Under the SIE
formulation, the modal expansion coefficient can be evaluated via
the inner product of two light fields over the nanostructures' surfaces
$\partial\Omega_{0}$ \citep{makitaloModesResonancesPlasmonic2014},
\begin{equation}
a^{(m,n)}=\frac{\iint_{\partial\Omega_{0}}\mathbf{x}_{L}^{(n)}\cdot\mathbf{b}^{(m)}\mathrm{d}S}{\iint_{\partial\Omega_{0}}\mathbf{x}_{L}^{(n)}\cdot\mathbf{x}^{(n)}\mathrm{d}S},
\end{equation}
where $\mathbf{x}_{L}^{(n)}$ and $\mathbf{x}^{(n)}$ are the left
and right eigenvectors associated with the $n$th eigenmode with an
eigenvalue $\lambda^{(n)}$, and $\mathbf{b}^{(m)}$ is the vector
representation of the focal field inversely designed from the $m$th
eigenmode and evaluated on the surface of the nanodisk(s) $\partial\Omega_{0}$.
In detail, the left eigenvector is defined by, in comparison with
Eq.~(\ref{eq:SIE_eigenmode_real}),
\begin{equation}
\mathbf{x}_{L}^{(n)}\left(\mathcal{Z}_{0}+\mathcal{Z}_{1}\right)=\lambda^{(n)}\mathbf{x}_{L}^{(n)},
\end{equation}
and the vector of the designed focal field is
\begin{equation}
\mathbf{b}^{(m)}=\left[\mu_{0}\mathbf{J}_{f}^{(m)},\epsilon_{0}\mathbf{M}_{f}^{(m)}\right],
\end{equation}
where $\mathbf{J}_{f}^{(m)}=\hat{n}_{0}\times\mathbf{H}_{f}^{(m)}$
and $\mathbf{M}_{f}^{(m)}=-\hat{n}_{0}\times\mathbf{E}_{f}^{(m)}$
are the electric and magnetic surface current densities arising from
the designed focal field over the surface $\partial\Omega_{0}$.

The eigenmode's near field is solved directly in Eq.~(\ref{eq:SIE_eigenmode_real})
over the surface $\partial\Omega_{0}$. The designed focal field on
$\partial\Omega_{0}$ can be evaluated via the surface double integral
in Eq.~(\ref{eq:efocal}), where the reference field $\mathbf{E}_{\infty}$
in the integrand is calculated through Eq.~(\ref{eq:eref_iFT}) by
fast Fourier transform (FFT) \citep{frigoDesignImplementationFFTW32005}.
As a result, the reference field $\mathbf{E}_{\infty}$ is a 2D matrix
on a square grid, and we need to use Eq.~(\ref{eq:efocal}) rather
than Eq.~(\ref{eq:efocal_FT}) for the focal field calculation on
$\partial\Omega_{0}$. Here, the double integral is implemented through
the 2D numerical integration based on the trapezoidal rule \citep{cheneyNumericalMathematicsComputing2007}.

The expansion coefficients for the nanodisk monomer, dimer, and tetramer
are visualized in three colormap matrices in Figs.~\ref{fig:exp_coeffs}(a),
(b), and (c), respectively. The $m$th focal field $\mathbf{E}_{f}^{(m)}$
is inversely designed from the corresponding $m$th eigenmode's near
field in the focal plane. For the monomer, the designed focal fields
$\mathbf{E}_{f}^{(5)}$, $\mathbf{E}_{f}^{(8)}$, $\mathbf{E}_{f}^{(9)}$,
and $\mathbf{E}_{f}^{(10)}$ largely overlap with the eigenmodes from
which they are designed, respectively. In other words, these eigenmodes
can be effectively mode-matched by tightly focused vector beams via
our inverse design approach. However, the designed focal fields $\mathbf{E}_{f}^{(m)}$
with $m=1,2,11,\text{ and }12$ do not effectively match the corresponding
desired eigenmodes and instead other eigenmodes are strongly excited.
The designed focal fields $\mathbf{E}_{f}^{(m)}$ with $m=3,4,6,\text{ and }7$
strongly, but not exclusively (since other eigenmodes are also excited),
excite the desired corresponding eigenmodes. In the nanodisk dimer
case, the effective mode-matching focal fields can be obtained for
the eigenmodes $\mathbf{E}^{(n)}$ with $n=1\text{ to }6$, but the
focal fields $\mathbf{E}_{f}^{(m)}$ with $m=7\text{ to }12$ jump
to near fields that effectively match with the eigenmodes $\mathbf{E}^{(3)}$,
$\mathbf{E}^{(4)}$, $\mathbf{E}^{(5)}$, $\mathbf{E}^{(6)}$, $\mathbf{E}^{(5)}$,
and $\mathbf{E}^{(6)}$, respectively. For the nanodisk tetramer,
the effective mode-matching focal fields are obtained for the eigenmodes
$\mathbf{E}^{(n)}$ with $n=2,3,5,6,7,8,\text{ and }9$. However,
the designed focal fields $\mathbf{E}_{f}^{(10)}$ and $\mathbf{E}_{f}^{(11)}$
will excite the eigenmodes $\mathbf{E}^{(8)}$ and $\mathbf{E}^{(9)}$,
respectively. The designed focal fields $\mathbf{E}_{f}^{(1)}$ and
$\mathbf{E}_{f}^{(12)}$ always excite two eigenmodes simultaneously,
including the one from which the individual focal field is inversely
designed.

\section{Conclusion and discussion\label{sec:conclusion}}

We have developed an inverse design approach of generating a focused
light field that potentially matches with an eigenmode's near field
in optical nanoantennas. We began with a rigorous analysis of eigenmodes
in the the optical nanoantennas of interest. In this work, a nanodisk
monomer, dimer, and tetramer were considered, and the first twelve
eigenmodes in each nanostructure were studied using our BEM eigenmode
solver that is based on the N-Müller formulation of the SIEs. In our
inverse design strategies, the eigenmode's near field in the nanodisk's
middle symmetric plane (which is chosen to coincide with the focal
plane) is set as the desired focal field, which is inversely propagated
to the pupil plane where the designed pupil field is obtained. In
this backward propagation, we took into account the beam-like characteristic
of the pupil field and the fact the pupil field is band-limited, as
well as the inherent dependencies of different field components. The
eigenmodes were categorized into four groups by the time-averaged
energies associated with the transverse and longitudinal electric
and magnetic field in the focal plane, and the dominant field components
were chosen for uniquely determining the beam-like pupil field. The
designed focal field's expansion coefficients into the first twelve
eigenmodes were evaluated. For several eigenmodes, effective and exclusive
mode excitations were achieved by designing the focal field accordingly
from the eigenmodes via our developed inverse design method. This
work can have a significant impact on optical switching, near- and
far-field engineering in nano-optics.

\section*{Acknowledgments}

The authors thank Martti Kauranen for valuable advice on improving
the manuscript. This work was financially supported by the Academy
of Finland {[}projects 308393 and 320166 (PREIN){]}.

\end{document}